\documentstyle[amssymb,aps]{revtex}

\begin{document}
\title{On the physical mechanisms of relaxation time distribution in disordered
dielectrics.}
\author{V.A.Stephanovich$^1$, M.D.Glinchuk$^2$, B.Hilczer$^3$, L.Jastrabik$^4$}
\address{$^1$Institute of Physics of Semiconductors, NASc of
Ukraine,\\
Pr.Nauki 45, 252650 Kiev, Ukraine \\
$^2$Institute for
Problems of Materials Science, NASc of Ukraine,\\
Krjijanovskogo
3, 252180 Kiev, Ukraine \\
$^3$Institute of Molecular Physics Polish Academy of Sciences,\\
Poznan, Poland,\\
$^4$Institute of Physics, Academy of
Sciences of the Czech Republic,\\
Na Slovance 2, 180 40 Praha 8,\\
Czech Republic }

\date{\today}
\maketitle

\begin{abstract}
The distribution function of relaxation times in disordered dielectrics has
been calculated in the random field theory framework. For this purpose, we
first consider the dynamics of single two-orientable impurity electric
dipole in a random electric field $E$ created by the rest of impurities in
disordered ferroelectric. This dynamics is conveniently described by
Langevin equation. Relaxation time $\tau $ is then a reciprocal probability
(calculated on the base of Fokker-Planck equation) of the dipole transition
through barrier in a double-well potential (corresponding to two possible
dipole orientations), distorted by a random fields.

The obtained dependence $\tau (E)$ made it possible to obtain the expression
for relaxation times distribution function $F(\tau )$(via random fields
distribution function $f(E)$. Latter function has been calculated
self-consistently in the random field theory framework. Nonlinear random
field contribution and effects of spatial correlations between impurities
have also been taken into account. It was shown that nonlinear contribution
of random field gives asymmetric shape of $F(\tau)$, while in linear case it
is symmetric.

Comparison of calculated $F(\tau)$ curves with those extracted from
empirical Cole-Cole (CC), Davidson-Cole (DC), Kohlrausch-William-Watts (KWW)
and Havriliak-Negami (HN) functions had shown, that they correspond to mixed
ferro-glass phase with coexistence of short and long-range order. Different
forms of $F(\tau )$ are determined by linear (CC) or nonlinear (DC, KWW, HN)
contributions of random field.
\end{abstract}

\section{Introduction}

The anomalies of dynamic properties are the characteristic feature of the
disordered ferroelectrics, polymers and composites. In particular, strong
frequency dispersion of dielectric or magnetic susceptibility was observed
in many dipole or spin glasses (see e.g. [1] and ref. therein). The
dispersion is commonly attributed to existence in the disordered systems of
broad spectrum of relaxation times which can be extracted from the observed
frequency dependence of susceptibility [2]. The background of such
phenomenological approach is the model of a superposition of Debye relaxors
with different relaxation times $\tau $. In such a model the dynamic
quantities like time or frequency dependence of polarization, system
susceptibility etc. can be calculated by averaging of single relaxor
characteristics with relaxation time distribution function $F(\tau )$ in
supposition of parallel (independent) relaxation processes. The key point of
such approach is the form of distribution function. Only a few simple
empirical forms were proposed (rather than calculated in some physical
model) for $F(\tau )$. One of them were considered by Fr\"ochlich [3] who
assumed that in arbitrary range $\left[ \tau _0,\tau _1\right] $ $F(\tau )$
has constant positive value and outside this range it equals zero. But such
form of $F(\tau )$ appeared unable to explain the experimental data for real
systems with dielectric response described by Cole-Cole (CC), Davidson-Cole
(DC), Havriliak-Negami (HN) and other more complex forms, than simple Debye
equation [4]. Different and complex enough empirical forms of $F(\tau )$
were shown to be necessary to describe non-Debye dielectric response [5].
However the extraction of $F(\tau )$ from the experimental data makes it
impossible to understand the physical reasons of these properties anomalies.

Recently the calculations of linear [6,7] and nonlinear [8] dynamic
dielectric response were performed for ferroelectric relaxors like PMN, PST,
PLZT. Random electric field produced by substitutional disorder and other
lattice imperfections (which are known to be the characteristic features of
all the disordered systems) was considered as the main reason of their
properties peculiarities via its influence on the barriers between a dipole
orientations. Vogel-Fulcher law in relaxation time temperature dependence,
stretch-exponential behaviour of dynamic polarization [6] and several
peculiarities of dc field non-linear susceptibility [8] were obtained.

In present work random field theory approach was applied for calculation of
relaxation time distribution function. The influence of random electric
field on the parabolic and rectangular barriers was considered. The
relaxation time distribution function was calculated with the help of random
fields distribution function allowing for linear [9] and non-linear random
fields contribution as well as spatial correlation effects in the impurity
subsystem [10].

\section{Connection between distribution functions of relaxation times and
random fields}

Random field is known to be the characteristic feature of the disordered
systems. The substitutional disorder, vacancies of ions, impurities and
other imperfections are the sources of random electric and elastic field in
the system. These random fields $E$ make relaxation times $\tau $ of the
dipoles to vary from point to point in a sample, i.e. $\tau =\tau (E)$.
Possible forms of $\tau (E)$ will be considered below. The distribution of
random electric and elastic fields was calculated earlier in [9], [10] and
[11] respectively in the statistical theory framework. The connection
between $\tau $ and $E$ permits to calculate the distribution function of $%
\tau $ for known distribution of $E$ by means of theory of probability. In
the case when $\tau (E)$ is single-valued (monotonous) function of $E$ the
theory of probability gives [12]:

\begin{equation}
F(\tau )=f(E(\tau ))\left| \frac{dE(\tau )}{d\tau }\right|  \label{dz1}
\end{equation}
Here $F(\tau )$ and $f(E)$ are the relaxation time and random field
distribution functions respectively, $E(\tau )$ is inverse function to $\tau
(E)$. The Eq.(\ref{dz1}) can be thought about as simple change of variables
in normalization integrals
\begin{equation}
\int_{\tau }F(\tau )d\tau =\int_{E}f(E)dE  \label{dz11a}
\end{equation}
where integration is performed over domains of $\tau $ and $E$ variation.
The only difference is the modulus of the derivative because the
distribution function has to be positive.

In more general case of non-monotonous behaviour of $\tau (E)$, when one $%
\tau $ value correspond to several $E$-values ($E_1,E_2...E_n$), the space
of $E$ values should be divided into $n$ regions (containing $E_1,E_2...E_n$
points), where function $\tau (E)$ is monotonous (see Fig.1). Now, for the
entire $E$-domain, $F(\tau )$ can be represented as a sum of expressions
like Eq.(\ref{dz1}) over regions $E_i(\tau )$ of monotonous behaviour of $%
\tau (E)$ [12]:

\begin{equation}
F(\tau )=\sum_{i=1}^n f(E_i(\tau ))\left| \frac{dE_i(\tau )}{d\tau }%
\right|  \label{dz2}
\end{equation}
Eq.(\ref{dz2}) is general expression for the distribution function
of one quantity via the distribution function of another one for a
given connection between them. In particular, Eq.(\ref{dz2}) can
be used for the calculation of random field distribution function
$f(E)$ allowing for nonlinear and spatial correlation effects via
that in the linear case. In general case internal random field
$E(r_i)$ can be written as:
\begin{equation}
E_\gamma (\overrightarrow{r_i})=\varepsilon _\gamma (\overrightarrow{r_i}%
)+\sum\limits_{m=2}^p\alpha _m\varepsilon _\gamma ^m(\overrightarrow{r_i})
\label{dz3}
\end{equation}

\begin{equation}
\varepsilon _\gamma (\overrightarrow{r_i})=\sum\limits_{k,j}(\varepsilon
_k)_\gamma (r_{ij})  \label{dz4}
\end{equation}
where $(\varepsilon _k)_\gamma $ is $\gamma $-component of the field
produced in the observation point $\overrightarrow{r_i}$ by a source of $k^{%
\text{th}}$ type (e.g. dipoles, point charges, dilatational centers)
situated at the point $\overrightarrow{r_j}$; $\alpha _m$ are the
coefficients of nonlinearity of $m^{\text{th}}$ order of the host lattice,
their dimension being the inverse electric field in $(m-1)$-th power.

In the linear case ($\alpha _m=0$) the random field distribution function $%
f_l$ can be calculated in the statistical theory framework, namely

\begin{equation}
f_l(\varepsilon )=\overline{\delta (\varepsilon -\varepsilon ^{\prime
}(r_{ij}))},  \label{dz5}
\end{equation}
where averaging (marked by bar) of $\delta $-function makes it possible to
count all the spatial configurations of random field sources $\varepsilon
^{\prime }(r_{ij})$ leading to definite random field value $\varepsilon
_\gamma \equiv \varepsilon $. Averaging, neglecting the correlations between
random field sources in Eq.(\ref{dz5}), leads to well-known expression of
statistical theory [13,14]:

\begin{equation}
f_l(\varepsilon )=\frac 1{2\pi }\int\limits_{-\infty }^\infty \exp
(i\varepsilon t-\sum\limits_kn_kF_k(t))dt  \label{dz6}
\end{equation}

\begin{equation}
F_k(t)=\int\limits_{-\infty }^\infty d^3r(1-\exp (it\varepsilon _k(r)))
\label{dz7}
\end{equation}
Here $n_k$ and $\varepsilon _k(r)$ are respectively the concentration and
electric field of $k^{\text{th}}$ random field source.

One can see, that the form of $F_k(t)$ and thus $f_l(\varepsilon)$ depends
strongly on the form of $\varepsilon _k(r)$. So, $f_l(\varepsilon)$ could
not only be of Gaussian form (which is realized, e.g., for point charges
with $\varepsilon _k(r)\sim 1/r^2$), but also of Lorentzian form (this is
the case for electric dipoles with $\varepsilon _k(r)\sim 1/r^3$).

In the disordered dielectrics the main source of random fields is impurity
electric dipole which can have several discrete orientations in a host
dielectric lattice (see Ref. [9] for details). In this case, which is of
interest for present consideration, the function $f_l(\varepsilon)$ chould
be calculated with additional quantum statistical averaging over
aforementioned discrete orientations. This was done self-consistently in
Ref. [9] (see also [10]) for particular case of two-orientable dipoles,
point charges and dilatational centers as the random field sources:

\begin{equation}
f_{l}(E)=\frac{1}{2\pi }\int\limits_{-\infty }^{\infty }\exp \left[
it(E-E_{0}L)-n_{1}B_{1}\left| t\right| -n_{2}B_{2}\left| t\right|
^{3/2}-n_{3}B_{3}t^{2}\right] dt  \label{dz8}
\end{equation}

\[
B_{1}=\frac{\Omega _{0}}{9}\frac{1+\nu }{1-\nu }p,B_{2}=\frac{32}{15}\left(
\frac{\pi Ze}{2\varepsilon _{0}}\right) ^{3/2},B_{3}=\frac{16\pi }{15}%
r_{c}^{3}\left( \frac{d^{\ast }}{\varepsilon _{0}r_{c}^{3}}\right) ^{2},\
L=\int\limits_{-\infty }^{\infty }f_{l}(E)\tanh \left( \beta E\right) dE.
\]
Here $L=\overline{\left\langle \left\langle d^{\ast }\right\rangle
\right\rangle }/d^{\ast }$ is long-range order parameter (the number of
coherently oriented impurity electric dipoles, see Ref. [9] for dtails), $%
E_{0}=4\pi (n_{3}d^{\ast 2})/\varepsilon _{0}$ is the mean value of random
electric field (in energy units), produced by the impurity dipoles, $d^{\ast
}=1/3d\gamma (\varepsilon _{0}-1)$ is the electric dipole moment, $\gamma $
and $\varepsilon _{0}$ are reaspectively the Lorentz factor and static
dielectric permittivity of the host lattice, $n_{1}$, $n_{2}$ and $n_{3}$
are respectively the concentrations of dilatational centers, point charges
and electric dipoles, $Ze$ and $\Omega _{0}$ are the point defect charge and
elastic moment, $p$ and $\nu $ are the host lattice piezoelectric component
and Poisson coefficient respectively.

Thus Eqs. (6), (7), (8) give the possibility to calculate the distribution
function of random field induced independently, i.e. without correlation, by
any number of random field sources.

Spatial correlations and nonlinear effects can be taken into account on the
base of Eq.(3) ($\alpha _{n}\neq 0$). For such more general case the random
field distribution function can be expressed through $f_{l}(\varepsilon )$
with the help of expression like Eq.(2) [15]:

\begin{equation}
f(E)=\sum\limits_{n}f_{l}(\varepsilon _{n}(E))\left| \frac{d\varepsilon
_{n}(E)}{dE}\right|  \label{dz9}
\end{equation}
where $\varepsilon _{n}$ are the real roots of the algebraic equation

\begin{equation}
E-\varepsilon -\alpha _{2}\varepsilon ^{2}-...-\alpha _{n}\varepsilon ^{n}=0
\label{dz10}
\end{equation}

We have to emphasize that even for simple enough (e.g. Gaussian, Lorentzian)
form of $f_{l}(\varepsilon )$, coefficient $\left| \frac{d\varepsilon _{n}(E)%
}{dE}\right| $ in Eq.(9) will be function of $E$ by virtue of Eq (\ref{dz10}%
). So, the nonlinear distribution function $f(E)$ will never be the simple
algebraic sum of $f_{l}(\varepsilon _{n})$. The shape of $f(E)$ was
calculated recently with the help of Eqs. (\ref{dz9}), (\ref{dz10}) for the
contribution of nonlinear terms of the second ($\alpha _{2}\neq 0$) and
third ($\alpha _{3}\neq 0$) order [15,16].

\section{Influence of electric field on relaxation time}

The ordinary ferroelectric materials like PbTiO$_{3}$, BaTiO$_{3}$ with
polar long-range order are known to have unique relaxation time
characterizing the rate of macroscopic polarization restoration after
external perturbation of the system.

Disordered systems like relaxors have broad spectrum of relaxation times,
because their most probable states are dipole glass (DG) with short-range
order polar clusters of $r_{c}$ size, imbeded into paraelectric phase and/or
mixed ferro-glass phase (FG) with coexistence of short- and long-range
order. Since an orientable electric dipole is usually the nucleation centre
of a cluster, the random reorientation of the dipoles can be the main
mechanism of the relaxation in the aforementioned disordered systems. It is
commonly belived that the probability of a dipole reorientation temperature
dependence is obeyed to Arrhenius law

\begin{equation}
\frac{1}{\tau }=\frac{1}{\tau _{0}}\exp \left( -\frac{U}{T}\right)
\label{dz11}
\end{equation}
where $U$ is a height of a barrier between equivalent dipole orientations.

To obtain the connection between relaxation time $\tau $ and random field $E$%
, let us consider the relaxational dynamics of single impurity (for
definiteness two-orientable) dipole in a random electric field, created by
the rest of impurity dipoles. It is clear, that in such relaxation process
the random field will be time dependent and will play a role of a random
force, exerted on the given impurity dipole. For such motion we can write
following Langevin equation (see, e.g. \cite{chalub})

\begin{equation}
\frac{dx}{dt}=-\frac{\partial U(x)}{\partial x}+f(t),  \label{dz12}
\end{equation}
where $x$ is the displacement of impurity ion (so that impurity dipole
moment $d=ex$, $e$ is ion charge), $U(x)$ determines the actual potential
energy of dipole (e.g. for two-orientable dipole $U(x)$ is a conventional
double-well potential) and $f(t)$ is a random force with known correlator.
Without loss of generality we can put
\begin{equation}
<f(t)>=0,\ <f(t)f(t^{\prime })>=D\delta (t-t^{\prime }),  \label{dz121}
\end{equation}
where $D$ is diffusion coefficient (see below). Langevin equation (\ref{dz12}%
) is nonlinear stochastic differential equation. It can be used, in
particular, to determine the probability for impurity dipole to transit from
shallower to deeper well of double-well potential (i.e. overbarrier
''bounce''), deformed by a random field (Fig. 2a). This problem is
equivalent to the problem of Brownian particles diffusion through barrier in
the potential $U(x)$ (see \cite{fizkin} and refs therein). For the
probability $n(x,t)$ that diffusing particle has coordinate $x$ at time $t$
the Fokker-Planck equation can be derived from (\ref{dz12}). For
two-orientable dipole (one-dimensional case) it reads

\begin{eqnarray}
&&\frac{\partial n}{\partial t}+\frac{\partial j}{\partial x}=0,
\label{dz13} \\
&&j=-D\frac{\partial n}{\partial x}-bn\frac{\partial U(x)}{\partial x}.\
\nonumber
\end{eqnarray}

Here $D$ is diffusion coefficient for particles from well $A$ to well $B$ of
potential $U(x)$ (Fig.2a). For it the Einstein relation holds
\[
D=bT,
\]
where $b$ is particle mobility, $\ j$ is diffusion flux. It is seen that $%
\partial n/\partial t=0$ in Eq. (\ref{dz13}) when
\begin{equation}
n=n_{eq}(x)=A\exp \left( -\frac{U(x)}{T}\right) ,  \label{dz131}
\end{equation}
i.e. when $n$ is equilibrium Maxwell-Boltzmann distribution.

It can be shown (see, e.g. \cite{fizkin} and refs therein) that probability
of transition from well $A$ to well $B$ (i.e. reciprocal relaxation time) is
ratio of stationary (at $\partial n/\partial t=0)$ flux $j_{0}$ and number
of particles $N_{A}$ in well $A$%
\begin{equation}
p=\tau ^{-1}=\frac{j_{0}}{N_{A}}.  \label{dz132}
\end{equation}
Calculation with respect to boundary condition $n(B)=0$ (reflecting the fact
that at $t=0$ all particles are in well $A$) yields

\begin{equation}
j=\frac{Dn(A)}{\int\limits_{A}^{B}\exp (U(x)/T)dx},\text{ }%
N_{A}=\int\limits_{A}n(A)\exp \left( \frac{U(x)}{T}\right) dx  \label{dz15}
\end{equation}

One can see from Eqs. (\ref{dz132}), (\ref{dz15}), that the relaxation time
depends mainly on the form of potential $U(x)$. It can be shown that the
main contribution to the integrals originates from potential $U(x)$ maximum
region. So, we can approximate $U(x)$ by parabola near this point and
integrate from $-\infty $ to $+\infty $ (this is standard steepest descent
method for integrals calculation). For zero random field ($E=0$) $U(x)\equiv
U_{0}(x)$ is symmetric with respect to $y$ axis and this procedure gives
standard Arrhenius law (\ref{dz11}) for relaxation time. For nonzero random
local (i.e. in the given impurity location) electric field the symmetric
form of $U_{0}(x)$ is distorted
\begin{equation}
U(x)=U_{0}(x)\pm eE_{loc}x,\ U_{0}=U-ax^{2}+bx^{4},\ U=\frac{a^{2}}{4b}.
\label{dz16}
\end{equation}
Here $U$ is barrier height and for undistorted potential we choose the
simplest polynomial approximation. In subsequent calculations we will call
this barrier shape ''parabolic''.

Consider now aforementioned potential $U_{0}(x)=U-ax^{2}+bx^{4},$ which has
maximum at $x_{1}=0$ and minima at $x_{2,3}=\pm \sqrt{a/2b}$. In the spirit
of steepest descent method we considered the approximation of $U(x)$ (\ref
{dz16}) near its maximum at $x=0$ for small fields $E$. This made it
possible to carry out the integration in Eqs. (15), (16) by the steepest
descent method, that yields

\begin{equation}
P=\frac{1}{\tau }=\frac{1}{\tau _{0}}\exp \left( -\frac{U+(eE_{loc})^{2}/4a}{%
T}\right)  \label{dz18}
\end{equation}
where $1/\tau _{0}$ is temperature and field independent combination
involving constant b and $U_{0}(x)$ parameters. Note, that at $E=0$ we once
more obtain Arrhenius law from (\ref{dz18}).

For rectangular potential of the form (see Fig.2b)

\begin{equation}
U_{0}(x)=\left\{
\begin{array}{c}
0,\ -x_{0}-\Delta \leq x\leq -x_{0} \\
U,\ -x_{0}\leq x\leq x_{0} \\
0,\ x_{0}\leq x\leq x_{0}+\Delta
\end{array}
\right.  \label{dz19}
\end{equation}
the integrals (\ref{dz132}), (\ref{dz15}) can be calculated exactly, which
yields

\begin{eqnarray}
P &=&\frac{\alpha ^{2}D\exp (-\alpha x_{0})}{\left[ 2/\alpha \exp (U/T)\sinh
(\alpha x_{0})+4/\alpha \sinh (\alpha \Delta /2)\cosh (\alpha x_{0}+\alpha
\Delta /2)\right] \left( \exp (\alpha \Delta )-1\right) },  \label{dz20} \\
\alpha &=&\frac{eE}{T},\ d=ex_{0}.  \nonumber
\end{eqnarray}

It is seen from the Fig. 2a that actually $\Delta \ll x_{0}.$ In this case
we can substantially simplify Eq (\ref{dz20}) and obtain

\begin{equation}
\frac{1}{\tau }=\frac{1}{\tau _{0}}\exp \left( -\frac{U\pm dE_{loc}}{T}%
\right)  \label{dz21}
\end{equation}
It is seen that at $E_{loc}=0$ we again arrive at Arrhenius law. Note that
steepest descent method gives also Eq.(\ref{dz21}).

One can see from Eqs. (\ref{dz18}), (\ref{dz21}), that electric field
decreases the barrier for dipole orientation along field direction and
increases it for opposite one. Thus, random electric field increases
relaxation time.

The final form of relaxation time dependence on $E$ can be obtained after
averaging Eq.(21) over possible orientations of electric dipole $d$.
Quantum-statistical averaging with the energy in the form $\Re =-dE_{loc}$
gives

\begin{equation}
\left\langle \tau \right\rangle =\overline{\tau _{0}}\frac{sp\exp (d^{\ast
}E/T-\Re /T)}{sp\exp (-\Re /T)}=\overline{\tau _{0}}\frac{ch(2d^{\ast }E/kT)%
}{ch(d^{\ast }E/kT)},\overline{\tau _{0}}=\tau _{0}\exp \left( \frac{U}{T}%
\right)  \label{dz22}
\end{equation}

In Eq. (\ref{dz22}) we substituted product $dE_{loc}$ by $d^{\ast }E$ where
effective dipole moment $d^{\ast }=(\varepsilon _{0}-1)\gamma d/3$ ($%
\varepsilon _{0}$ and $\gamma $ are respectively host lattice permittivity
and Lorentz factor).

Note, that such type of electric field influence on the barrier height was
supposed by us earlier (see e.g. [6-8] and ref. therein).

In what follows we will perform the calculations for rectangular barriers on
the base Eq.(\ref{dz22}), because the influence of electric field on
parabolic barrier (\ref{dz16}) is qualitively the same.

\section{Relaxation time distribution function}

\subsection{General equations}

Eq.(2) with respect to Eq.(22) makes it possible to calculate relaxation
time distribution function. In particular Eq.(22) leads to the following
connection between electric field and relaxation time

\begin{equation}
E(t)=kT\arccos h\left( \frac{1}{4}(t+\sqrt{t^{2}+8})\right) \equiv kTf_{0
\pm} (t);t\equiv \frac{\tau }{\overline{\tau _{0}}};t\geq 1  \label{dz23}
\end{equation}
where sighs "$\pm $" correspond to two branches of the function

\begin{equation}
\arccos h(x)=\ln (x\pm \sqrt{x^{2}-1})  \label{24}
\end{equation}

Allowing for these two branches as $i=1,2$ in Eq.(2) and substituting $E(t)$
for $E$ in Eq. (\ref{dz8}) (linear case) or in Eqs.(\ref{dz9}), (\ref{dz10})
(nonlinear case) one can obtain general expressions for relaxation times
distribution function. Since these expressions appeared to be quite
cumbersome, we considered the case when only electric dipoles are the
sources of random electric field i.e. $n_{1}=n_{2}=0$ in Eq.(\ref{dz8}). In
this case Eq.(\ref{dz8}) yields

\begin{equation}
f_{l}(E)=\frac{1}{\sqrt{2\pi }\Delta }\exp \left( -\frac{(E-E_{0}L)^{2}}{%
2\Delta ^{2}}\right) ,\ \Delta =\sqrt{2n_{3}B_{3}},  \label{dz25}
\end{equation}
i.e. linear random field distribution function $f_{l}(E)$ has Gaussian form.
Eq.(25) with respect to Eqs (\ref{dz2}) and (\ref{dz23}) leads to the
following expression of relaxation time distribution function
\begin{mathletters}
\begin{equation}
F(t)=\frac{b_1}{\sqrt{\pi }}\left\{ \left| \frac{df_{0+}}{dt}\right| \exp %
\left[ -b_2(vf_{0+}(t)-L(v,z))^2\right] +\left| \frac{df_{0-}}{dt}\right|
\exp \left[ -b_2(vf_{0-}(t)-L(v,z))^2\right] \right\}  \label{dz26a}
\end{equation}

\begin{equation}
\frac{df_{0\pm }}{dt}=\pm \frac 1{\sqrt{2}}\frac{1+t/\sqrt{t^2+8}}{\sqrt{%
t^2+t\sqrt{t^2+8}-4}}  \label{dz26b}
\end{equation}
where we introduced the dimensionless parameters suitable for numerical
calculations

\end{mathletters}
\begin{eqnarray}
b_1 &=&\frac{kT}{2\sqrt{n_3B_3}}\equiv vb_2;b_2=\frac{E_0}{2\sqrt{n_3B_3}}=%
\frac 12\sqrt{15\pi z};  \label{dz27} \\
z &=&n_3r_c^3;v=\frac{kT}{E_0}  \nonumber
\end{eqnarray}

Let us check the normalization of $F(t)$. The normalization condition for $%
f(E)$

\begin{equation}
\int\limits_{-\infty }^{\infty }f(E)dE=1  \label{dz28}
\end{equation}
transforms for $F(t)$ into contour integral, where contour runs first from $%
\infty $ to 1 over $t$ axis and then back - from $1$ to $\infty $ (see Fig.
3).

To perform integration we should pass from integration over $dt$ to the
integration over $df_{0}$ (with respect to $f_{0\pm }(1)=0$, $f_{0+}(\infty
)=\infty $, $f_{0-}(\infty )=-\infty $). This yields

\begin{eqnarray}
\frac{b_1}{\sqrt{\pi }}\left[ \int\limits_0^\infty \exp \left[
-(b_1f_0-b_2L(v,z))^2\right] df_0+\int\limits_{-\infty }^0\exp \left[
-(b_1f_0-b_2L(v,z))^2\right] df_0\right] &=&  \label{dz29} \\
\frac{b_1}{\sqrt{\pi }}\int\limits_{-\infty }^\infty \exp \left[
-(b_1f_0-b_2L(v,z))^2\right] df_0 &=&1  \nonumber
\end{eqnarray}

Therefore Eqs.(\ref{dz26a}), (\ref{dz26b}) represent normalized distribution
function of relaxation time. Any other integral involving function $F(t)$
may also be evaluated by this procedure.

One can see that $F(t)$ depends on random field sources concentration and
temperature via dimensionless parameters $z$, $v$ and order parameter $%
L(z,v) $. The dependence $L(z,v)$ actually determines the phase diagram of
the system under consideration. We calculated it earlier both for linear [9]
and nonlinear [10] random field contributions. It was shown, that $L=0$ for $%
n_{3}<n_{3c}$, $T<T_{g}$, where $n_{3c}$ and $T_{g}$ are respectively
critical concentration and freezing temperature for dipole glass (DG) state;
$L=1$ for $z\rightarrow \infty $ (mean field approximation) and $T\ll
E_{0}=kT_{cmf}$, where $T_{cmf}$ is ferroelectric (FE) phase transition
temperature. In intermediate range of concentrations ($\infty >n_{3}>n_{3c}$%
) only the part of dipoles is ordered, i.e. $0<L<1$ and ferro-glass (FG)
phase with coexistence of long and short range order appears.

Thus, Eqs.(\ref{dz26a}), (\ref{dz26b}) permit to calculate the distribution
function of relaxation times for DG, FG, FE and paraelectric phases via $%
L(z,v)$ dependence in these phases.

\subsection{Relaxation time distribution function in mean field approximation%
}

In the ferroelectric and paraelectric phases all properties of disordered
dielectrics, including distribution of relaxation times, can be calculated
in mean field approximation. In this approximation random field distribution
function has the form of $\delta $-function, i.e. $f_{mf}(E)=\delta
(E-E_{0}L)$ (linear case) and $\delta (E-E_{0}L(1+\alpha
_{3}E_{0}^{2}L^{2})) $ (nonlinear case [10]) with $L\neq 0$ for FE phase and
$L=0$ for paraelectric phase. Substitution of these functions into Eq.(1)
gives e.g. for linear case

\begin{equation}
F_{mf}(t)=\delta (E(t)-E_0L)\left| \frac{dE(t)}{dt}\right|  \label{dz30}
\end{equation}

Taking into account that $f(x)\delta (x-a)=f(a)\delta (x-a)$ and $\delta
(f(x))=\sum\limits_{k}\delta (x-x_{k})/\left| f^{\prime }(x_{k})\right| $
(see, e.g.,[18]), where $x_{k}$ are the real roots of equation $f(x_{k})=0,$
one can rewrite Eq.(\ref{dz30}) in the form

\begin{equation}
F_{mf}(t)=\delta (t-t_{mf})  \label{dz31}
\end{equation}
where $t_{mf}$ was shown to be the unique root of the equation $%
E(t)-E_{0}L=0 $ (linear case) or $E(t)-E_{0}L(1+\alpha _{3}E_{0}^{2}L^{2})=0$
(nonlinear case), see Fig.3. One can see from Eq. (\ref{dz23}) that in
paraelectric phase ($L=0$) $t_{mf}=1$, i.e. $\tau _{mf}=\overline{\tau _{0}}%
=\tau _{0}\exp (U/kT)$, whereas in FE phase ($L\neq 0$) additional
temperature dependence of relaxation time appears. In particular for $%
LE_{0}/kT\gg 1$ $\tau _{mf}=\tau _{0}\exp ((U+LE_{0})/kT)$ (linear case) or $%
\tau _{mf}=\tau _{0}\exp [(U+LE_{0}(1+\alpha _{3}E_{0}^{2}L^{2}))/kT]$
(nonlinear case), because of mean field influence on barrier height.

\subsection{Numerical calculations of relaxation time distribution function}

The calculation of relaxation time distribution function beyond of mean
field approximation were performed numerically on the base of Eq.(\ref{dz26a}%
) (linear case) and on the base of Eqs. (\ref{dz2}), (\ref{dz9}), (\ref{dz10}%
), (\ref{dz23}) (nonlinear case). Since the spectrum of relaxation times in
disordered systems is very broad ($1\leq \tau /\overline{\tau _{0}}=t<\infty
)$, we, following literature on this subject, plotted $F(t)$ in logarithnic
scale in $t$. By the same reasom we did not include $df_{0\pm }/dt$ $\sim
d(\ln t)/dt$ into our plots. Also, $f_{l}(E)$ was always taken in Gaussian
form (\ref{dz25}).

The results of calculations of relaxation time distribution function in the
linear case are represented in Fig.4. To exclude a paraelectric phase we
choose $v=T/T_{cmf}<1$ and considered two values of dipole concentrations $%
z=nr_{c}^{3}=1$ and 10 (curves 1 and 2 respectively). Due to the values of
parameters chosen, curve 1 corresponds to $F(t)$ for FG phase whereas curve
2 illustrates the transformation of $F(t)$ into the form close to $\delta $%
-function, which is realized as $z$ increases (system approaches to FE
phase). The shape of curve 1 reflects the Gaussian form (\ref{dz25}) of $%
f_{l}(E)$ . For non-Gaussian shape of $f_{l}(E)$ (e.g. Lorentzian,
Holtzmarkian etc, corresponding to different random field sources), the
shape of $F(t)$ in linear case will also reflect them. Since all
aforementioned curves are known to be symmetric functions, the linear
relaxation time distribution function also must be symmetric.

Let us proceed to calculation of $F(t)$ in nonlinear case. Since the
solutions of Eq. (\ref{dz10}) depend on the lattice symmetry we considered a
host lattice with a centre of inversion in the paraelectric phase, that is
characteristic for many disordered systems. In such a case only the terms
with odd powers of $\varepsilon $ remain in Eq.(\ref{dz10}). In our
calculations we retained only the first nonlinear term in Eq.(\ref{dz10}),
i.e. $\alpha _{3}\varepsilon ^{3}$. The sign and value of nonlinearity
coefficient $\alpha _{3}\equiv \alpha _{0}E_{0}^{2}$ appeared to influence
strongly the shape of $F(t)$ (see Fig.5 for $\alpha _{0}>0$ and Fig.6 for $%
\alpha _{0}<0$). One can see that for $\alpha _{0}>0$ $F(t)$ broadens and
shifts towards larger $t$ with $\alpha _{0}$ increase. Such behaviour can be
the result of barrier increase due to nonlinear random field contribution.
Shape of $F(t)$ transforms from almost symmetric to slightly asymmetric with
$\alpha _{0}$ increase. For example, for $\alpha _{0}=1$ the right ''wing''
of the curve is just Gaussian while left one decays faster then Gaussian.
Strongly asymmetric shape of $F(t)$ is peculiar feature for $\alpha _{0}<0$,
the line asymmetry being larger at larger $\left| \alpha _{0}\right| $ (see
Fig.6). The narrowing of $F(t)$ and its maximum position shift to the
smaller $t$ with $\left| \alpha _{0}\right| $ increase can be the result of
the barrier decrease. More essential asymmetry of $F(t)$ for $\alpha _{0}<0$
in comparison with $\alpha _{0}>0$ is the result of presence of only one
root of Eq. (\ref{dz10}) (and symmetric shape of $f(E)$) for $\alpha _{0}>0,$
whereas for $\alpha _{0}<0$ there are three different real roots of Eq. (\ref
{dz10}), which breaks $f(E)$ symmetric shape (see [15]). The curves depicted
in Figs. 5 and 6 correspond to FG phase with coexistence of short- and
long-range order. Calculations of $F(t)$ for DG state (where only
short-range ordered clusters exist ) lead to very broad curve with very slow
decay even in $\ln (t)$ scale.

\section{Discussion}

The distribution of relaxation times is usually extracted from observed
frequency dependence of dielectric susceptibility. This dependence for
ordered systems is known to be described by Debye law with the \ unique
relaxation time. To describe the observed dynamic susceptibility of the
disordered ferroelectrics, polymers and composites, several empirical
functions were proposed as of Debye law generalizations (see e.g. [4]).
Among them the most known are following:

\begin{equation}
\frac{\varepsilon ^{\ast }(\omega )-\varepsilon _{\infty }}{\varepsilon
_{0}-\varepsilon _{\infty }}=\left\{
\begin{array}{c}
\left( 1+(i\omega \tau _{CC})^{1-\kappa }\right) ^{-1}\ \ \ \ \ (a) \\
\left( 1+i\omega \tau _{DC}\right) ^{-\beta }\ \ \ \ \ \ \ \ \ \ \ \ \ (b)
\\
\left( 1+\left( i\omega \tau _{HN}\right) ^{\gamma }\right) ^{-\delta }\ \ \
\ \ \ \ \ (c)
\end{array}
\right.   \label{dz32}
\end{equation}
Eq.(\ref{dz32}) (a), (b), (c) represent, respectively, Cole-Cole (CC) ($%
0\leq \kappa <1$), Davidson-Cole (DC) ($0<\beta \leq 1$) and
Havriliak-Negami (HN) ($\gamma \leq 1$, $\delta \leq 1$) functions.

All these functions are written in the frequency domain, whereas in time
domain Debye relaxation is usually generalizes to Kohlrausch-Williams-Watts
(KWW) relaxation function, which is also currently called a {\em ''stretched
exponential''}

\begin{equation}
\Phi (t)=\exp \left( -t/\tau _{WW}\right) ^\alpha ,0<\alpha \leq 1
\label{dz33}
\end{equation}

Since there is no analytical expression for Eq. (\ref{dz33}) in frequency
domain, the numerical calculation of its Fourier transform was performed
[4]. It was shown that at $\gamma \delta =\alpha ^{1,2,3}$ the results are
identical to those for H-N function.

Empirical laws (\ref{dz32}), (\ref{dz33}) were applied for many years for
the description of slow relaxation processes in conventional glasses,
polymers, composites, disordered ferroelectrics etc. The data obtained by
several experimental techniques, including dielectric spectroscopy, nuclear
magnetic resonance, quasielastic neutron scattering , kinetic reactions etc.
were sucsessfully fitted by these laws (see e.g. [4], [19]). It was supposed
, that physical origin of the laws (\ref{dz32}) and (\ref{dz33}) was some
distribution of relaxation times $F(\tau ),$ so that
\begin{mathletters}
\begin{equation}
\frac{\varepsilon ^{\ast }(\omega )-\varepsilon _{\infty }}{\varepsilon
_{0}-\varepsilon _{\infty }}=\int\limits_{0}^{\infty }\frac{1}{1+i\omega
\tau }F(\tau )d(\ln \tau )  \label{dz34a}
\end{equation}

\begin{equation}
\exp \left( -t/\tau _{WW}\right) ^\alpha =\int\limits_0^\infty \exp \left(
-t/\tau _{WW}\right) F(\tau )d(\ln \tau )  \label{34b}
\end{equation}
The expressions (\ref{dz32}), (\ref{dz34a}),(\ref{dz34b}) made it possible
to extract relaxation time distribution function for all the empirical laws.
The results of such extraction, obtained in Ref. [5] are represented in
Fig.7 for Debye CC ($\varkappa =0,2$), DC ($\beta =0,6$), KWW ($\alpha =0,42$%
). Due to aforementioned relation between KWW and HN laws the shape of $%
F(\tau )$ for HN is similar to that for KWW function (see Fig.7). The
obtained distribution functions are empirical ones with strongly different
shapes. Namely, $F(\ln \tau )$ can be symmetric (CC), asymmetric (KWW, HN)
and strongly asymmetric (DC) function. To the best of our knowledge, the
physical mechanisms of such behaviour is not known out up to now.

To clarify the physical mechanisms let us compare the empirical curves from
Fig.7 with the curves in Figs. 4, 5, 6, obtained on the base of random field
distribution function. As it was shown in Section 4 the symmetric form of $%
F(\tau )$ is pertinent to linear approximation what is valid for small
enough random field sources concentrations. With increase of random field
sources concentration nonlinear and correlation effects become essential.
They result in $F(\ln \tau /\overline{\tau _{0}})$ small asymmetry for $%
\alpha _{0}>0$ (Fig.5) and strong asymmetry for $\alpha _{0}<0$ (Fig.6). The
comparison of the curves in Figs. 6 and 7 shows that the shape of the curves
2 and 3 in Fig.6 looks like that for KWW and DC laws respectively (Fig.7).
Relative position of their maxima is in conformity with that for curves 2
and 3 in Fig.6. The coinsidence of $\tau _{CC}$ and Debye relaxation time $%
\tau _{D}$ (see Fig.7) can be expected also from calculated curves in Fig.4
because with increase of random electric dipoles concentration curve 2 has
to be transformed into $\delta $-function (see Section 4). Since calculated
curves in Figs. 4, 5, 6, correspond to FG phase with coexistence of short-
and long-range order, we can conclude that this coexistence as well as small
(CC law) or large nonlinear effect contribution with negative coefficient
(KWW, DC laws) are the physical background of the considered empirical laws.
Note, that nonlinear and correlation effects with negative nonlinearity
coefficient decrease the system order, i.e. they makes it ''more
disordered'', as it was shown recently [10]. Under such conditions the role
of random fields and thus the distribution of relaxation times becomes more
essential in the peculiarities of physical properties of disordered
dielectrics.

\section{Conclusion}

The coincidence between the relaxation time distribution function obtained
from empirical laws and calculated on the base of random fields distribution
function gives evidence that random fields are indeed the main physical
reason for the distribution of relaxation times. Variety of distribution
function forms is due to "degree of disorder" (i.e. presence of different
amounts of different random field sources) in the systems under
consideration.

Suggested general formalism for relaxation times distribution function
calculation has pretty general nature (because of generality of statistical
method, see \cite{kniga,stoneh}) within assumptions made. It can be easily
generalized to other disordered systems like spin glasses.


\begin{figure}[tbp]
\caption{Schematic plot of nonmonotonous behaviour of $\tau (E)$
function. One value $\tau _0$ corresponds to four $E$ values ($E_1
- E_4 $). In the regions 1-4 function $\tau (E)$ is monotonous (it
 decreases in regions 1 and 3 and increases in 2 and 4). }
\end{figure}
\begin{figure}[tbp]
\caption{(a) - Parabolic barrier in the external electric field
$E$. It is seen that barrier height at $E \neq 0$ is larger than
that at $E=0$. Parabolic approximation for steepest descent method
is also shown.\\
(b) - Rectangular barrier at $E=0$ (solid line), $E<0$ (dashed line) and $%
E>0$ (dotted line). }
\end{figure}
\begin{figure}[tbp]
\caption{Schematic plot of dependence $E(t)$ (\ref{dz23}). Arrows
show the integration contour in (\ref{dz29}). We also show the
unique root $t_{mf}$ of equation $E(t)-E_{mf}=0$ (see Eqs
(\ref{dz32}), (\ref{dz33})), $E_{mf}=E_0L$ (linear case) or
$E_{mf}=E_0L(1+\alpha _0L^2)$. For the sake of illustration
$E_{mf}$ is plotted only for the case $L>0$ and $\alpha _0 >0$.}
\end{figure}

\begin{figure}[tbp]
\caption{Relaxation time distribution function for small random field
contribution - linear case. Parameters of calculations: $v=0,3$, $z=1$
(curve 1), $z=10$ (curve 2). }
\end{figure}

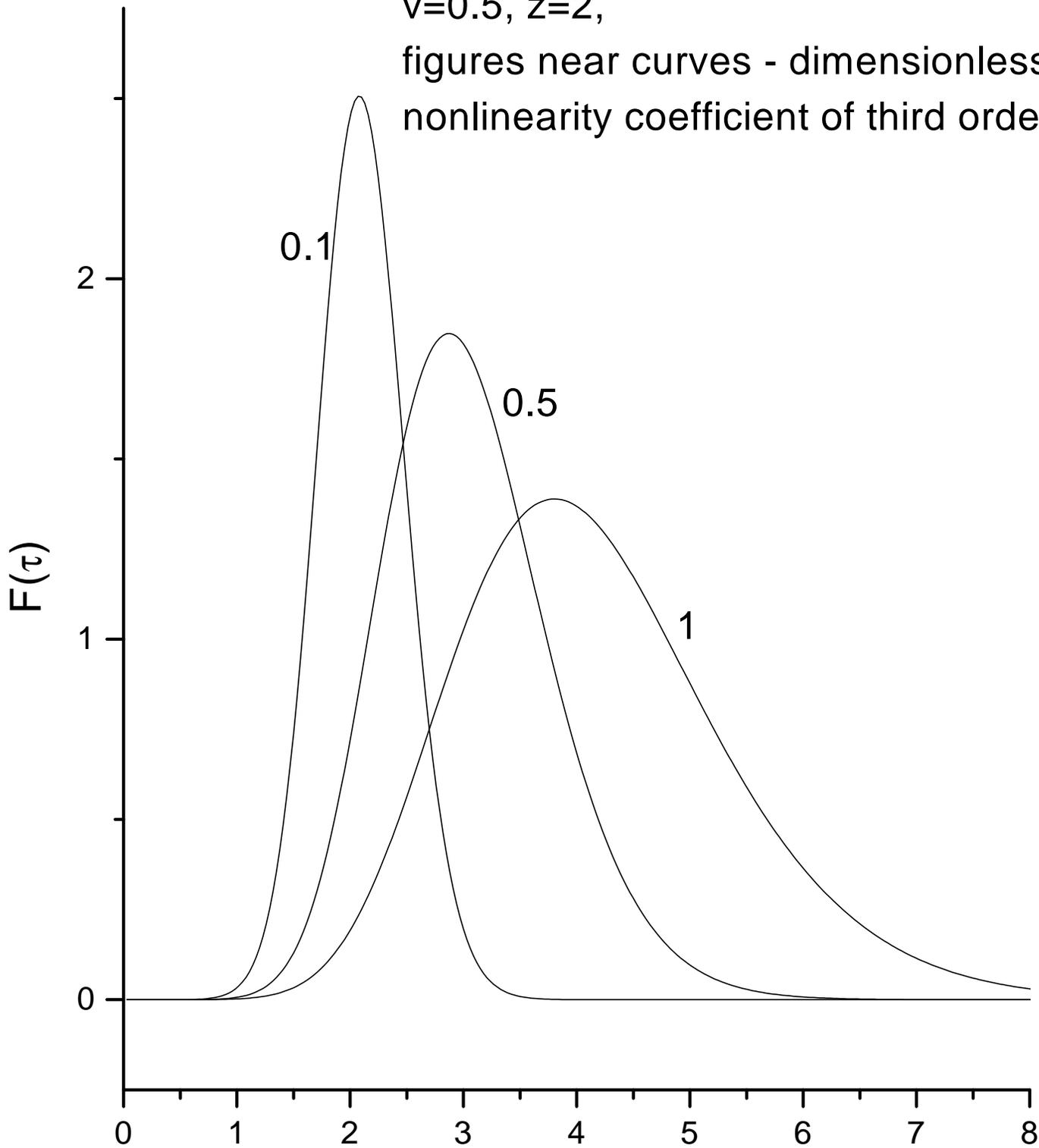
\begin{figure}[tbp]
\caption{Relaxation time distribution function for large random field
contribution - nonlinear case, with positive (a) and negative (b)
coefficient of nonlinearity. Parameters of calculations: $v=0,5$; $z=2$; $%
\protect\alpha _0=1; 0,5; 0,1$ (curves 1, 2, 3 respectively in (a)); $v=0,3$%
; $z=1$; $\protect\alpha _0=-0,01; -0,1; -0,3$ (curves 1, 2, 3 respectively
in (b)). }
\end{figure}

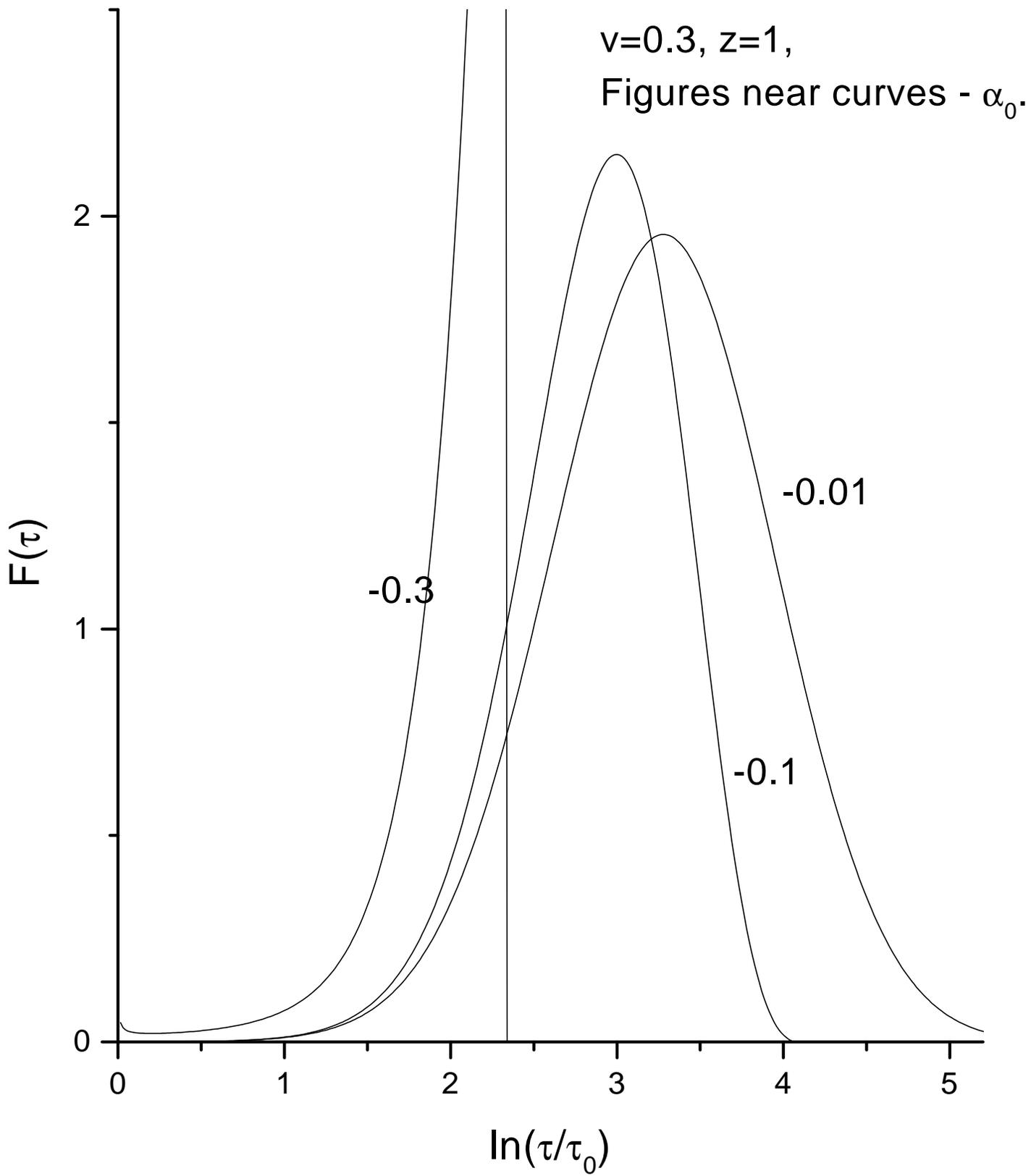
\begin{figure}[tbp]
\caption{Relaxation time distribution function for Debye (D), Cole-Cole
(CC), Davidson-Cole (DC) and Kohlrausch-Williams-Watts (WW) laws [5]. }
\end{figure}
\end{mathletters}

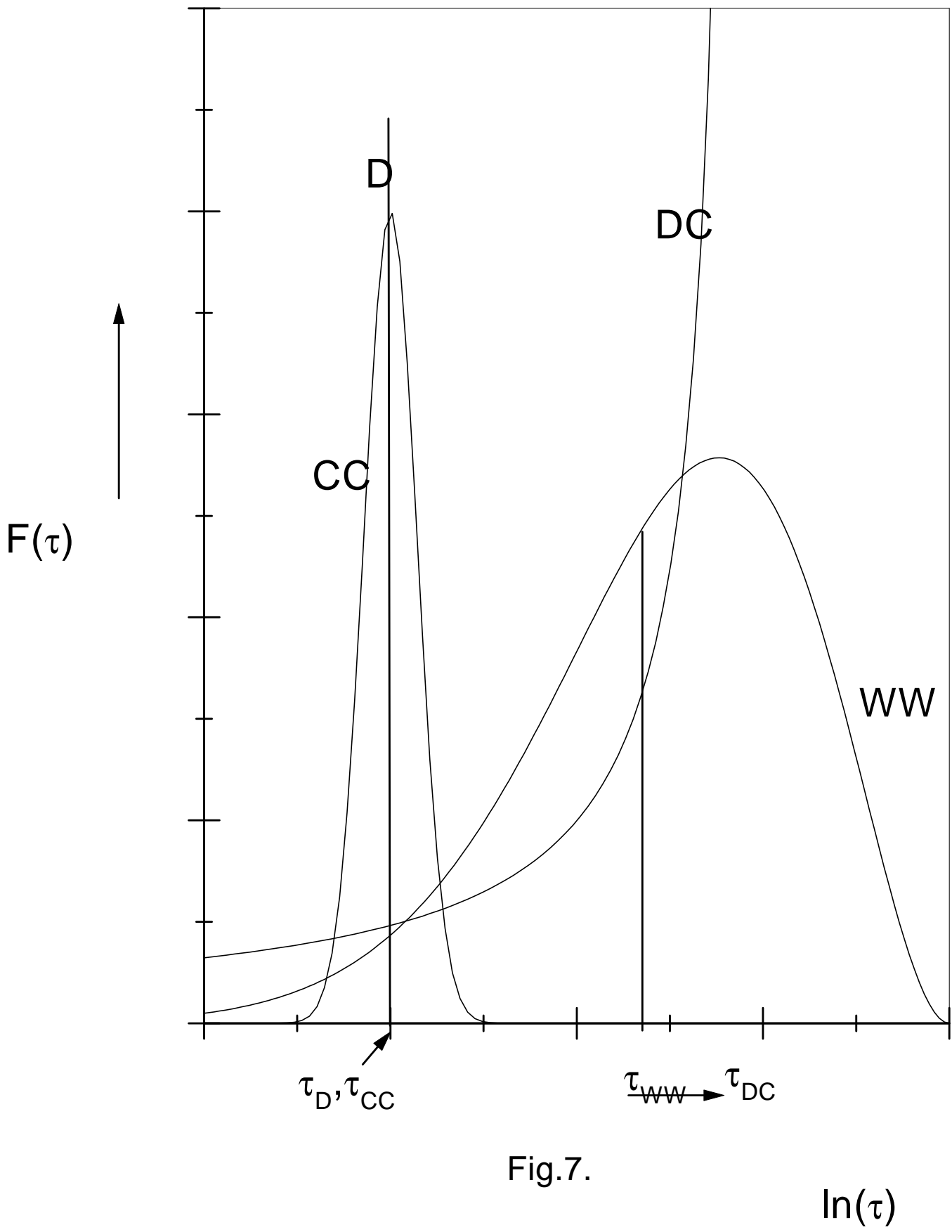
\end{document}